\documentclass[]{aastex631}

\newcommand{\msol}{$M_{\odot}$}

\defcitealias{Levan23}{L23}
\defcitealias{Darbha19}{D19}

\begin{document}

\title{Ultra-deep cover: an exotic and jetted tidal disruption event candidate disguised as a gamma-ray burst}

\correspondingauthor{R. A. J. Eyles-Ferris}
\email{raje1@leicester.ac.uk}

\author[0000-0002-8775-2365]{R. A. J. Eyles-Ferris}
\affiliation{School of Physics and Astronomy, University of Leicester, University Road, Leicester, LE1 7RH, UK}

\author[0000-0002-2137-4146]{C. J. Nixon}
\affiliation{School of Physics and Astronomy, Sir William Henry Bragg Building, University of Leeds, Woodhouse Ln., Leeds, LS2 9JT, UK}

\author[0000-0003-3765-6401]{E. R. Coughlin}
\affiliation{Department of Physics, Syracuse University, Syracuse, NY 13210, USA}

\author[0000-0002-5128-1899]{P. T. O'Brien}
\affiliation{School of Physics and Astronomy, University of Leicester, University Road, Leicester, LE1 7RH, UK}

\begin{abstract}
Gamma-ray bursts (GRBs) are traditionally classified as either short GRBs with durations $\lesssim 2$\,s that are powered by compact object mergers, or long GRBs with durations $\gtrsim 2$\,s that powered by the deaths of massive stars. Recent results, however, have challenged this dichotomy and suggest that there exists a population of merger-driven long bursts. One such example, GRB 191019A, has a $t_{90} \approx 64$\,s but many of its other properties -- including its host galaxy, afterglow luminosity and lack of associated supernova -- are more consistent with a short GRB. Here we propose an alternative interpretation: that GRB 191019A (which is located in the nucleus of its host) is an atypical jetted tidal disruption event (TDE). In particular, we suggest the short timescale and rapid decline, not expected for standard TDEs, are the result of an ``ultra-deep'' encounter, in which the star came well within the tidal radius of the black hole and promptly self-intersected, circularised, accreted, and launched a relativistic outflow. This model reproduces the timescale and luminosity through a prompt super-Eddington accretion phase and accounts for the lack of late optical emission. This would make GRB 191019A only the fifth jetted TDE and the first discovered ultra-deep TDE. The ultra-deep TDE model can be distinguished from merger-driven long GRBs via the soft X-ray flash that results from prompt self-intersection of the debris stream; the detection of this flash will be possible with wide-field and soft-X-ray satellites such as {\it Einstein Probe} or {\it SVOM}.
\end{abstract}


\section{Introduction} \label{sec:intro}

Gamma-ray bursts (GRBs) release vast amounts of energy, power relativistic jets, and are observable to cosmological distances (see, e.g., \citealt{woosley06, piran04, kumar15} for reviews). There are two main types of GRB, traditionally characterised by their $t_{90}$ (the time over which 90\% of the gamma-rays are detected): long GRBs with $t_{90}>2$\,s and softer gamma-ray spectra, and short GRBs with $t_{90}<2$\,s and harder spectra. This dichotomy self-consistently arises from the interpretation that long GRBs are powered by the deaths of massive stars, while short GRBs are powered by the merger of compact objects \citep{abbott17a,abbott17b,kasen17}.

However, recent studies have identified the possible existence of a population of GRBs that do not fit into this simple picture \citep{Gompertz23,Troja23,Levan23b}. One such candidate is GRB 191019A \citep[][hereafter \citetalias{Levan23}]{Levan23}, detected by the \textit{Neil Gehrels Swift Observatory} (\textit{Swift}). Superficially, GRB 191019A is a fairly typical long GRB with $t_{90} = 64.4 \pm 4.5$\,s at a redshift of $z=0.248$. However, when compared to the general long GRB population, it has a somewhat under-luminous afterglow and there is no evidence of later supernova emission as would be expected \citep[e.g.,][]{Woosley95}. In addition, its host galaxy has much less star formation than typical long GRB hosts. Together, these properties are more similar to those of short GRBs and drive \citetalias{Levan23}'s interpretation of this as a merger-driven long GRB, i.e., a GRB that was powered by the merger of two compact objects, but some unconstrained physical conditions led to the longer-than-expected release of photons.

What sets GRB 191019A apart from the rest of the merger-driven long GRB sample, however, is its location in the nucleus of its host galaxy consistent to $\lesssim 100$ pc. \citetalias{Levan23} suggest that the dense nuclear environment is naturally inclined to dynamical interactions that might produce these GRBs. The environment will also have a significant impact on the behaviour of the GRB, and one possibility is that the merger that powered GRB 191019A took place within the disk of an active galactic nucleus \citep[AGN,][]{Lazzati23}. The high-density circumburst medium could stretch the prompt emission from $\sim1$\,s to the observed time and result in two peaks in the prompt light curve, consistent with the behaviour of GRB 191019A. As noted by \citeauthor{Lazzati23} and \citetalias{Levan23}, however, there is little evidence for an AGN in the host galaxy. Late time X-ray observations limit any activity to a low luminosity and the infrared colours and optical spectrum point more towards a spiral or possibly a luminous infrared galaxy classification \citep{Wright10}.

Another possibility, which we explore here, is that GRB 191019A is a tidal disruption event (TDE), in which a star is torn apart by the central supermassive black hole (SMBH) and a fraction of the tidally disrupted debris is accreted \citep{rees88,gezari21}. The proximity of the event to the nucleus of the host galaxy is naturally explained in this picture, and the properties of GRB 191019A's host closely resemble the population of TDE hosts \citep{DeckerFrench20,Hammerstein21,Hammerstein23}. However, for a standard TDE (comprising the disruption of a main-sequence star by a black hole of mass $\sim 10^5-10^8 M_\odot$), the tidal debris returns to the SMBH on timescales of order months \citep[largely independent of the properties of the star being disrupted; e.g.,][]{bandopadhyay24}. The returning material forms an accretion flow that provides the observable emission, resulting in an electromagnetic transient that fades over timescales that are too long to be consistent with the $t_{90}$ duration of GRB 191019A. It is this discrepancy in timescales that led \citetalias{Levan23} to favour the merger-driven long GRB interpretation.

However, it is possible that some, more exotic, TDEs could undergo much faster evolution and therefore be compatible with the observed behaviour of GRB 191019A. In a standard TDE, the star approaches the SMBH with a pericentre distance that is close to, but smaller than, the tidal radius at which the star is completely destroyed. When the pericentre distance is much smaller than the tidal radius, the star suffers substantial in-plane distortion and vertical compression near pericentre, and this is the ``deep'' regime studied by, e.g., \citet{carter82, carter83, bicknell83, laguna93, stone13, evans15, tejeda17, steinberg19, norman21, coughlin22b, kundu22} and \cite{nixon22}. As the pericenter continues to decrease, the star is eventually directly captured by the black hole, particularly where the SMBH is high mass, and no observable emission is produced. However, prior to this limit being reached, there is the possibility of a third regime -- that which we denote the ``ultra-deep'' limit. In this limit, analyzed by \citet{evans15} and \citet[][hereafter \citetalias{Darbha19}]{Darbha19}, the in-plane stretching of the star is so severe prior to pericenter, that the star is able to wrap around the black hole and promptly self-intersect. This rapid interaction of the debris at the pericentre of the stellar orbit provides an opportunity for much shorter duration emission than is produced in typical TDEs.

In this letter, we examine the possibility that the emission observed in GRB 191019A is produced by an ultra-deep TDE. In Section 2, we discuss the properties of ultra-deep TDEs. In Section 3, we calculate the timescales for the emission processes that result from this model and compare them to GRB 191019A. In Section 4 we discuss our results and conclude.

\section{The ultra-deep TDE model}
\label{sec:model}
Stars are destroyed by the tidal field of an SMBH when they reach a sufficiently small distance from the SMBH. This distance, known as the tidal radius and typically denoted $r_{\rm t}$, is estimated by equating the tidal field of the SMBH to the surface self-gravitational field of the star \citep{Hills75}, resulting in the canonical expression $r_{\rm t} \simeq R_{\star} (M_{\bullet}/M_\star)^{1/3}$ (see \citealt{coughlin22c} for a more accurate, analytical method of determining the tidal radius for different types of stars). The impact parameter of the stellar orbit is then defined as the ratio of the tidal radius to the star's pericentre distance, $\beta \equiv r_t / r_p$, and ultra-deep TDEs have $\beta_{\rm c} < \beta < \beta_{\rm dc}$, where $\beta_{\rm dc}$ is the maximum impact parameter that can be achieved without directly capturing the star, and $\beta_{\rm c}$ is the critical value required for prompt self-intersection; as we shall see below, regions of parameter space that facilitate ultra-deep TDEs have $\beta_{\rm dc} > \beta_{\rm c} \gg 1$. During the star's ingress (i.e., after crossing the tidal radius but prior to reaching pericenter), the tidal field of the SMBH stretches the star approximately in the direction from the star to the black hole, and \citetalias{Darbha19} show that at the time the star's centre of mass reaches $r_p$, its radial extent is given by
\begin{equation}
    R_p = R_{\star} \left(\frac{8}{5}\beta^{1/2}+\frac{2}{5}\beta^{-2}\right) \simeq  \frac{8}{5}R_{\star}\beta^{1/2},
    \label{eq:R_p}
\end{equation}
where the last equality assumes $\beta \gg 1$ (see also \citealt{stone13} for the $\beta^{1/2}$ scaling). An ultra-deep TDE can occur\footnote{This is a necessary condition; the relativistic advance of periapsis angle also needs to be $\gtrsim 2\pi$.} when $\beta$ is sufficiently large that the tidally deformed and stretched star intersects itself, i.e., if $R_p \geq 2\pi r_p$. The critical $\beta$ for ultra-deep TDEs, $\beta_c$, is therefore (see Equation~8 in D19)
\begin{equation}
    \beta_c \simeq \left(\frac{5\pi}{4}\right)^{2/3}\left(\frac{M_{\bullet}}{M_\star}\right)^{2/9}.
    \label{eq:beta_c}
\end{equation}
For the star to be sufficiently stretched by the time it reaches pericentre to generate prompt self-intersection we require $\beta > \beta_{\rm c}$. Stars with pericentre inside the direct capture radius $r_{\rm dc}$ will typically be swallowed whole, and so we also require that $\beta_{\rm c} < \beta_{\rm dc}$, where $\beta_{\rm dc} = r_{\rm t}/r_{\rm dc}$ is the maximum $\beta$ that can be achieved without directly capturing the star. This leads to the constraint 
\begin{equation}
\label{eq:mconstraint}
\frac{M_{\bullet}}{M_\star} < \chi^{-9/8}\left(\frac{4}{5\pi}\right)^{3/4}\left(\frac{R_\star c^2}{GM_\star}\right)^{9/8}\,,
\end{equation}
where $\chi = r_{\rm dc}/r_{\rm g}$ with $r_{\rm g} = GM_{\bullet}/c^2$. The value of $\chi$ depends on the spin of the black hole and the inclination of the orbit, ranging from $\chi=1$ for maximally spinning and prograde encounters to $\chi = (1+\sqrt{2})^2 \approx 5.83$ for maximally spinning and retrograde encounters; for a Schwarzschild black hole $\chi=4$ \citep[e.g.][]{will12}. Therefore, the constraint on the black hole mass for low-mass main sequence stars with $M_\star/M_\odot \simeq R_\star/R_\odot$ to be able to achieve prompt self-intersection of the stellar debris is $M_{\bullet}/M_\star \lesssim 1.2\times 10^5$ for a maximally spinning black hole and a retrograde encounter, $M_{\bullet}/M_\star \lesssim 1.8\times 10^5$ for a Schwarzschild black hole, and $M_{\bullet}/M_\star \lesssim 8.7\times 10^5$ for a maximally spinning black hole and a prograde encounter.

We therefore expect that the ultra-deep TDE model is most applicable to low mass SMBHs, with masses of order $10^5 M_\odot$, for which $\beta_{\rm c} \approx 30$ for a Sun-like star and $r_p \approx 7r_{\rm g}$. \citetalias{Levan23} used the host galaxy's stellar mass to estimate the SMBH mass as a few $10^7 M_\odot$. If this is correct, then it is unlikely that an ultra-deep TDE is possible in this system. However, by using SMBH masses determined via modelling of the TDE lightcurve, \citet{Hammerstein23} find that, contrary to the correlations used by \citetalias{Levan23} to estimate the SMBH mass, there is no significant correlation between stellar mass and SMBH mass for TDE hosts when using lightcurve modeling (see \citealt{mockler19}). A significantly lower SMBH mass cannot, therefore, be ruled out.

We note that the above equations are conservative on a number of grounds. For example, \citetalias{Darbha19} compare the predictions of the above equations with the results of simulations (both geodesic orbits and hydrodynamical simulations) and found that $\beta_{\rm c}$ for self-intersection could be significantly smaller.  Additionally, \citetalias{Darbha19} note that the equations above are derived using Newtonian gravity, and the GR tidal stretching of the star is stronger than Newtonian gravity predicts. It may, therefore, be possible that in relativistic gravity and for specific types of stars, the deformation at pericentre is sufficient to lead to self-intersection shocks for higher black hole masses. This will need testing in the future with dedicated numerical simulations.

Equations \ref{eq:beta_c} and \ref{eq:mconstraint} are also specific to the case when the entirety of the star encircles the black hole, leading to the self-intersection of the bulk of the stellar debris. It is also possible, specifically when the star has a core-envelope structure in which the former is considerably denser than the latter, for the low-density envelope of a star to be stripped at significantly larger radii than is predicted by $r_{\rm t}$ (which uses the average stellar density). In such scenarios, a relatively small fraction of the debris can encircle the SMBH and produce a prompt self-intersection and -- because of the disparity in densities -- a comparatively weaker shock. The simulations performed by \cite{evans15} -- who found prompt self-intersections for values of $\beta$ that are much more modest than those predicted by Equation \ref{eq:beta_c} -- were in this regime, as they used a $\gamma = 4/3$ polytrope to model their star, which has a much more centrally concentrated core (and extended envelope) than the $\gamma = 5/3$ polytrope used in \citetalias{Darbha19}.

For ultra-deep TDEs there are 3 key phases in the evolution: (1) the self-intersection shock, (2) the prompt accretion flow comprising the shocked material which we refer to as the ``prompt accretion emission,'' and (3) accretion of material that is much more weakly bound to the black hole and returns on much longer timescales, which we refer to as the ``delayed accretion emission.'' In the next section we analyze these three phases in detail and compare the predicted properties to the observations of GRB 191019A.

\section{GRB 191019A as an ultra-deep TDE}
\label{sec:analysis}

\subsection{Stream self-intersection}
We first discuss the emission from the self-intersection shock. Following \citetalias{Darbha19}, the shock generated during the immediate self-intersection deposits thermal energy of order $\rho v_s^2 \simeq 3M_\star v_s^2/4\pi R_\star^3$ into length $l_c\sim R_\star$ of the debris stream where $v_s$ is the shock velocity. The shock velocity is approximately equal to the pericentre velocity, $v_s \sim v_p \sim (GM_{\bullet}/r_p)^{1/2}$. The optical depth of the emitting surface layer is given by $\tau=\rho\kappa\Delta R$ where $\kappa = 0.4$ cm$^{2}$ g$^{-1}$ is the opacity and $\Delta R$ is its thickness. \citetalias{Darbha19} show that the thickness of the surface layer is given by $\Delta R \lesssim (cR_\star/\rho \kappa v_p)^{1/2}$ and the thermal energy in the surface layer is therefore given by
\begin{equation}
    \Delta E \sim 2\pi R_\star l_c \Delta R \rho v_s^2 \sim 2\pi l_c \left(R_\star^3  v_p^3 c\rho / \kappa\right)^{1/2}
    \label{eq:shock_energy}
\end{equation}
Assuming $M_{\bullet} = 10^{5.5}$ \msol\,and $\beta=\beta_c=41.5$, this implies an energy of $\sim 6.9 \times 10^{48}$ erg. This energy is released over a timescale comparable to the light crossing time, $t_{\rm lc} \sim R_\star / c$. For the Solar type star, $t_{\rm lc} \sim 2.3$ s and the luminosity is therefore $\sim 3.0\times10^{48}$ erg s$^{-1}$. Assuming the shock emission to be suitably represented by a single blackbody of radius $R_\star$, the effective temperature is $\sim 3.1$ keV.

This energy is well below that of \textit{Swift}'s Burst Alert Telescope (BAT), and this in addition to the short timescale means that \textit{Swift} would only be able to detect such emission if its X-Ray Telescope (XRT) was fortuitously aligned with the transient. Unfortunately, in the case of GRB 191019A, \textit{Swift} was unable to slew for 52 minutes and therefore this thermal emission was not detectable. To confirm this, we downloaded the XRT spectra from the UK \textit{Swift} Science Data Centre\footnote{\url{https://www.swift.ac.uk/}} and re-examined them using \texttt{XSPEC v12.13.0} \citep{xspec}. However, we identified no significant evidence of a softer thermal component. Wide-field X-ray instruments, such as those carried by \textit{Einstein Probe} \citep{Yuan22} or the \textit{Space Variable Objects Monitor} (\textit{SVOM}), stand a much better chance of detecting this emission which could be used to differentiate between GRBs and ultra-deep TDEs. Assuming a Planck cosmology \citep{Planck18}, \textit{Einstein Probe}'s Widefield X-ray Telescopes would be sensitive to the predicted emission in GRB 191019A's case to a redshift of $\lesssim0.8$ although the positional accuracy would not be sufficient to source the emission to a galactic nucleus.

\subsection{Prompt accretion emission}
\label{sec:prompt_accretion}

The prompt accretion will release an enormous amount of energy over a period comparable to the viscous timescale of the system. \citetalias{Darbha19} calculate the viscous timescale of the disc by making some assumptions about the disc structure (e.g. that the disc semi-thickness is of order the radius of the star). However, as the accretion rate is highly super-Eddington, we need to use equations that account for this. We therefore use those derived by \citet{Strubbe09} for a super-Eddington flow. In this model, the disc angular semi-thickness, $H/R$ is
\begin{equation}
    \frac{H}{R} = \frac{3}{4}f\left(\frac{10 \dot{M}}{\dot{M_{\rm Edd}}}\right)\left(\frac{R}{R_S}\right)^{-1}\left[\frac{1}{2} + \left(\frac{1}{4}+\frac{3}{2}f\left(\frac{10 \dot{M}}{\dot{M_{\rm Edd}}}\right)^{2}\left(\frac{R}{R_S}\right)^{-2}\right)^{1/2}\right]^{-1}
    \label{eq:H_R}
\end{equation}
where $f=1-(R_{\rm S}/2R)^{1/2}$ for a maximally-rotating, prograde SMBH and $f=1-(3R_S/R)^{1/2}$ for a non-rotating SMBH; $\dot{M}$ is the accretion rate; $\dot{M}_{\rm Edd}$ is the Eddington rate; and $R_{\rm S}=2GM/c^2$ is the Schwarzschild radius. The viscous timescale at radius $R$ is then
\begin{equation}
    t_{\rm visc} \simeq \alpha^{-1} \left(\frac{H}{R}\right)^{-2} \left(\frac{R^3}{G M_{\bullet}}\right)^{1/2}
    \label{eq:viscous_timescale}
\end{equation}
where $\alpha$ is the dimensionless accretion disk viscosity parameter \citep{Shakura76}, typically found to be $\approx 0.2-0.4$ for fully ionised disks \citep[e.g.][]{martin19}.

GRB 191019A's isotropic equivalent energy is $E_{\rm iso} \sim 10^{51}$ erg and we assume a jet opening angle of 10\degr\,and a radiative efficiency of $\eta\sim 0.1$. This implies a total energy $\sim3.8\times10^{\rm48}$ erg. Again we assume $M_{\bullet} = 10^{5.5}$ \msol and over GRB 191019A's $\sim50$ s rest frame timescale, this yields an Eddington ratio of $\sim1850$. Such a highly super-Eddington event provides suitable conditions for the launch of a relativistic jet. This jet results in the observed gamma ray emission and causes a broadband afterglow as it collides with the surrounding medium in the same way as a typical GRB.

\begin{figure}
\plotone{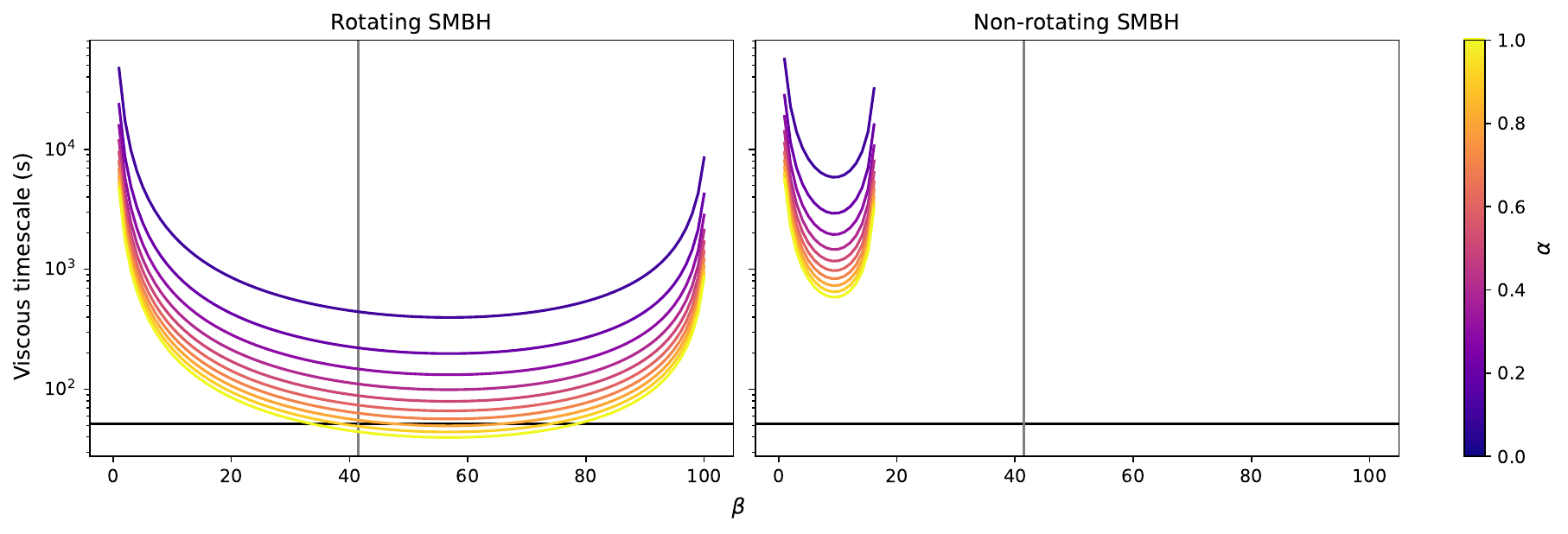}
\caption{The dependence of the viscous timescale on $\beta$ and $\alpha$ for rotating (left) and non-rotating $10^{5.5}$ \msol\,SMBHs (right).  We show to $r_p = r_{\rm dc}$ and $\alpha$ is varied from 0.3 to 1 in steps of 0.1.  In each panel, the black horizontal line indicates GRB 191019A's rest frame $t_{90}$ and the grey vertical line indicates $\beta_c=41.5$ for a $10^{5.5}$ \msol\, SMBH
\label{fig:viscous_timescale}}
\end{figure}

The assumptions made above result in a viscous timescale $\sim$ a few times too large to be consistent with the behaviour of GRB 191019A. However, there are systems where $\alpha$ has been observed to be significantly greater than 0.3. For instance, the disks around Be stars have been found to have $\alpha>0.5$, particularly during disk build-up and at smaller radii \citep{Rimulo18,Granada21}. In some cases, values of $\alpha\sim1$ have even been inferred \citep{Carciofi12}. The ionisation structure of the disk, primarily driven by radiation from the central star, has been shown to have a significant impact on $\alpha$. In the case of GRB 191019A, the enormous Eddington ratio and the extreme physics it invokes could produce a sufficiently strong radiation field and lead to high values of $\alpha$. Alternatively, it is possible that the self-intersection shock could provide such a field. We therefore vary $\alpha$ and find that $\alpha\gtrsim0.8$ and a rotating SMBH can reproduce the required timescale, as shown in Figure \ref{fig:viscous_timescale}.

\begin{figure}
\plotone{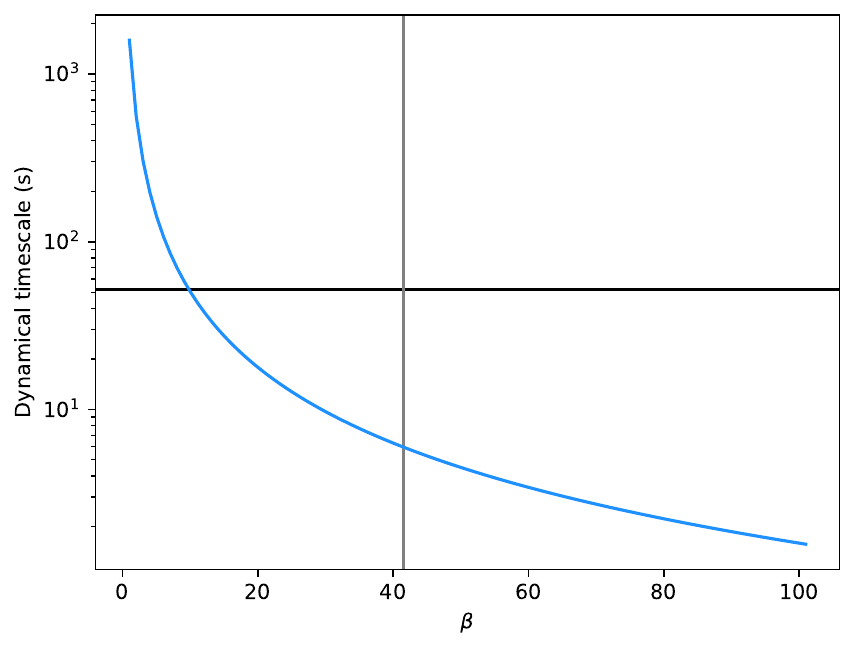}
\caption{The dependence of the dynamical timescale on $\beta$. We show to $r_p = r_{\rm dc}$. The black horizontal line indicates GRB 191019A's rest frame $t_{90}$ and the grey vertical line indicates $\beta_c=41.5$ for a $10^{5.5}$ \msol\, SMBH. 
\label{fig:dynamical_timescale}}
\end{figure}

Alternatively, and possibly more likely, is that this ``hyper-Eddington'' regime could affect the nature of the accretion flow itself; rather than forming a disk, the flow may form an essentially spherical structure. Similar (quasi-)spherical accretion models have been previously proposed to explain the behaviour of highly super-Eddington TDEs featuring relativistic jets \citep[e.g.][]{Coughlin14,Dai18}. In this model, Equation \ref{eq:H_R} would not apply and $H/R\sim1$ with no dependence on the rotation of the SMBH. The relevant timescale is therefore not the viscous timescale, but rather the much shorter dynamical timescale
\begin{equation}
   t_{\rm dyn} =  \left(\frac{R^3}{GM_{\bullet}}\right)^{1/2}
   \label{eq:t_dyn}
\end{equation}
and in Figure \ref{fig:dynamical_timescale}, we show that a spherical accretion model can readily achieve the required timescale.

An important consideration for a quasi-spherical accretion flow is the collimation of the GRB. In many quasi-spherical accretion models \citep[e.g.][]{Coughlin14,Dai18}, it is predicted that an outflow propagates along a centrifugally supported funnel and is collimated into a narrow relativistic jet. However, the ability of the gas to collimate the outflow is intricately tied to the run of pressure with spherical radius within the accretion flow, with more rapidly declining pressure profiles providing less collimation (e.g., \citealt{bromberg07, zakamska08, kohler12}). In the event that the outflow cannot be successfully collimated, the energy injection could instead result in a wide-angle and relativistically expanding outflow, somewhat akin to the `dirty fireball' model \citep[e.g.][]{Rhoads03}. Assuming such an expansion to be quasi-spherical, the total energy of the event would be comparable to the $E_{\rm iso}\sim10^{51}$ erg found by \citetalias{Levan23}. This leads to an estimated Eddington ratio of $\sim5\times10^5$ and an accretion rate of $\sim8\times10^4$ \msol yr$^{-1}$. This accretion rate is remarkably consistent with the predictions of \citet[][Fig 2]{evans15} whose simulations also predict an inflated accretion flow, albeit not spherical and with an accretion timescale of hours rather than seconds.

\subsection{Delayed accretion}
We have seen above that the material that is able to promptly self-intersect forms an accretion flow that evolves on very short timescales compared to a standard TDE. However, it is unlikely that the entire stellar debris is involved in this early phase and, instead, some of the debris escapes the central regions on a range of orbits encompassing the marginally bound orbit. The bound portion of this material will return at later times to the black hole and power late emission that is on timescales more typical of a standard TDE. The amount of mass contained in the bound portion is uncertain, but for the low-mass SMBH considered here, the return rate of material may still be comparable to or somewhat in excess of Eddington (e.g., \citealt{wu18} show that the fallback rate from a standard TDE onto a $10^{5.5}M_{\odot}$ SMBH can reach $\sim 10^{2}\dot{M}_{\rm Edd}$). If we assume the peak rate is comparable to the Eddington rate of the SMBH, then from the calculations presented in \citet{EylesFerris22}, the peak $g$ band magnitude for GRB 191019A (at $z=0.248$) is $\gtrsim 28.2$. This is well below the limits found by \citetalias{Levan23} and therefore their non-detection is consistent with our model of an ultra-deep TDE.

\section{Discussion}
\label{sec:discussion}

We have shown that the ultra-deep TDE model of \citetalias{Darbha19} is capable of reproducing the behaviour observed in GRB 191019A: the viscous timescale is sufficiently short for the prompt accretion to occur over a timescale comparable to the $t_{90}$ of the GRB, and its super-Eddington nature could power a relativistic jet to produce the observed emission. This would make GRB 191019A only the fifth detected jetted TDE\footnote{At time of writing.} and the first belonging to the ultra-deep class. The lack of later detections by \citetalias{Levan23} is also consistent with this model.

An ultra-deep TDE requires the stellar pericenter distance to be well within the tidal radius, as this yields the substantial stretching of the star prior to reaching pericenter that is necessary for prompt self-intersection, but outside of the direct capture radius. As discussed in \citetalias{Darbha19} and in Section \ref{sec:model} here, the values of $\beta$ ($=r_{\rm t}/r_{\rm p}$, with $r_{\rm t}$ the canonical tidal radius and $r_{\rm p}$ the pericenter distance) necessary to achieve prompt self-intersection are on the order of tens. Stars that are scattered into this region of angular momentum space must therefore be in the pinhole regime, where individual two-body scatterings yield changes in the specific angular momentum that are of the order the specific angular momentum itself (in contrast to the diffusive regime, where stars approach the $\beta \simeq 1$ limit over multiple orbital times; see, e.g., \citealt{frank76, lightman77, cohn78, merritt13, stone16}). 

In general we expect the likelihood of a star reaching these pericenter distances to be small relative to the likelihood of reaching smaller $\beta$'s (or, in some cases, being directly captured), purely as a consequence of the geometrical requirements of being placed onto an orbit with a narrow range of pericenters. The probability of being in the ultra-deep regime can be determined if the distribution of specific angular momenta (including the projection onto the spin axis of the SMBH) at large distances is known or assumed, and in general is an integral of the probability distribution function of the stellar angular momentum over the region in angular momentum space that yields pericenter distances outside of the direct capture radius but within $\beta_{\rm c}$, where $\beta_{\rm c}$ is given (approximately) by Equation \ref{eq:beta_c}. If the specific angular momenta of stars scattered into the loss cone are isotropically distributed (this is almost certainly a very good approximation, given that the distances from which stars tend to be scattered into the loss cone are many orders of magnitude larger than the tidal radius; e.g., \citealt{stone16}), then the integral can be calculated using the formalism described in \citet{coughlin22}\footnote{One can also use a Monte Carlo approach as in, e.g. \citet{kesden12}, but the use of the single integral described in \citet{coughlin22} is computationally cheap and enables the efficient exploration of the dependence of the probabilities on the parameters $\beta_{\rm c}$ (which is just a proxy for black hole mass for a given stellar mass) and black hole spin.}.

\begin{table}
\centering
    \begin{tabular}{|c|c|c|c|c|}
    \hline
     & $a=0$ & $a=0.5$ & $a=0.9$ & $a=0.999$ \\
  %   \hline
  %  $\beta_{\rm c} = 11.6$ ($M_{\bullet} = 10^{3}M_{\odot}$) & 8.52\% & 8.53\% & 8.54\% & 8.54\% \\ 
  %  \hline
   %  $\beta_{\rm c} = 19.3$ ($M_{\bullet} = 10^{4}M_{\odot}$) & 4.60\% & 4.61\% & 4.65\% & 4.67\% \\ 
    \hline
    $\beta_{\rm c} = 24.9$ ($M_{\bullet} = 10^{4.5}M_{\odot}$) & 2.78\% & 2.81\% & 2.90\% & 2.95\% \\ 
    \hline
    $\beta_{\rm c} = 32.1$ ($M_{\bullet} = 10^{5}M_{\odot}$) & 0.742 \% & 0.810\% & 1.01\% & 1.13\% \\
     \hline
    $\beta_{\rm c} = 36.5$ ($M_{\bullet} = 10^{5.25}M_{\odot}$) & $0.00206$\% & 0.0818\% & 0.323\% & 0.487\% \\
    \hline
    $\beta_{\rm c} = 41.5$ ($M_{\bullet} = 10^{5.5}M_{\odot}$) & 0 & 0 & 0.0412\% & 0.185\% \\
     \hline
    $\beta_{\rm c} = 47.2$ ($M_{\bullet} = 10^{5.75}M_{\odot}$) & 0 & 0 & 0 & 0.0421\% \\
    \hline
    \end{tabular}
    \caption{The probability of having a TDE with $\beta > \beta_{\rm c}$, where $\beta_{\rm c}$ is the approximate $\beta$ required for prompt self-intersection, for the SMBH masses, $M_{\bullet}$, in the left column and the SMBH spins, $a$, in the top row; here $\beta_{\rm c}$ is calculated assuming the star has a solar mass and radius. In cases where the probability is zero, $\beta_{\rm c}$ requires a pericenter distance that is inside the direct capture radius for a prograde orbit (i.e., all stars are directly captured and there are no observable TDEs).}
    \label{tab1}
\end{table}

Table \ref{tab1} gives the probabilities that result from this formalism for the $\beta_{\rm c}$ shown in the left column, with the corresponding SMBH masses shown in parentheses that arise from assuming the progenitor has a solar mass and radius (i.e., $R_{\star} = 1R_{\odot}$ and $M_{\star} = 1M_{\odot}$ in Equation \ref{eq:beta_c}), and the SMBH spins shown in the top row. These numbers therefore represent the relative number of stars that will be scattered into the loss cone and will have a point of closest approach that is within $r_{\rm t}/\beta_{\rm c}$, but will not be captured by the SMBH. Entries with a value of precisely zero require a pericenter distance that is within the direct capture radius of a prograde orbit, i.e., all stars will be directly captured by the SMBH, irrespective of SMBH spin and the orientation of the stellar orbit with respect to that spin. 

This table shows that the probability of achieving these orbits is always on the order of $1\%$ or less\footnote{As the SMBH mass decreases, the limiting $\beta$ is eventually not determined by $\beta_{\rm c}$, but instead by the requirement that the periapsis advance angle be $\gtrsim 2\pi$. This is important for the top-most row, where the pericenter distance with $\beta_{\rm c}$ is $\sim 19 GM_{\bullet}/c^2$, and at which the periapsis advance angle is substantially smaller than $2\pi$. Therefore, for a $10^{4.5}M_{\odot}$ SMBH, the probabilities are smaller than those reported in Table \ref{tab1}; for example, if we require that the pericenter distance of the star be $< 6GM_{\bullet}/c^2$, then the probability with $a = 0$ ($a = 0.999$) is $0.38$\% ($0.56$\%).}. We also see that the probability is always less than the Newtonian prediction of $1/\beta_{\rm c}$, as the Newtonian limit assumes that stars can penetrate to a pericenter distance of zero without being directly captured, and that SMBHs with masses $\gtrsim 10^{5.25} M_{\odot}$ must have a high spin in order to produce any ultra-deep TDEs. The number of ultra-deep TDEs we would expect per galaxy and per unit time is the product of the values given in Table \ref{tab1} and the loss-cone filling rate of $\sim 10^{-5} - 10^{-4}$ gal$^{-1}$ yr$^{1}$. As only a small fraction of stars are likely to be in the pinhole regime, the corresponding rate is $\lesssim 10^{-7}$ gal$^{-1}$ yr$^{-1}$. It is important to note, however, that the observed rate of ultra-deep TDEs is convolved with the underlying distribution of SMBH masses (and spins) at the low-mass end, which is observationally not well constrained -- some observational estimates find a roughly flat distribution (in black hole mass, not the log thereof) of black hole masses at the low-mass end (e.g., \citealt{graham07, davis14,mutlu16}), while other observational (e.g., \citealt{marconi04, shankar04}) and theoretical (e.g., \citealt{hopkins07, merloni08, shankar09, volonteri10}) works find a flat distribution in the log of the black hole mass and hence a diverging number of black holes with small masses. The steep power-law decline of the observed TDE luminosity function (e.g., \citealt{vanvelzen18, lin22, charalampopoulos23, yao23, guolo23}) is consistent with the former (i.e., a flat distribution in $M_{\bullet}$) if TDEs are powered by fallback accretion \citep{coughlin23}.

While the precise rate of ultra-deep TDEs is difficult to constrain without a better understanding of the SMBH mass function, it is clear from Table \ref{tab1} that they are, in general, rare. It is therefore not surprising, and indeed broadly consistent with this model, that of the (large) number of GRBs that have been observed to date, GRB 191019A appears almost unique. \citet{Troja23} estimate the rate of merger-driven long GRBs to be $0.04 - 0.8$ Gpc$^{-3}$ yr$^{-1}$ which implies a rate of order $10^{-8} - 10^{-7}$ gal$^{-1}$ yr$^{-1}$ with nuclear GRBs on the lower end, consistent with the expectations for ultra-deep TDEs described above.

GRB 191019A is an outlier in the merger-driven, long GRB sample. This is for two reasons -- its location in the nucleus of the host galaxy and no confirmed accompanying kilonova \citep[see][for the evidence for kilonovae in other merger-driven, long GRBs]{Troja23,Gompertz23,Levan23b}. However, the UVOIR upper limits obtained by \citetalias{Levan23} are relatively shallow relative to expectations for a kilonova at that time and therefore a merger-driven model remains consistent. \citetalias{Levan23} also point to GRB 050219A as a GRB with strikingly similar properties, and host galaxy properties, to GRB 191019A \citep{Tagliaferri05,Rossi14} and which also lacks confirmation of a kilonova, but is not located close to the nucleus of its host. While an offset from a galactic nucleus does not immediately rule out the ultra-deep TDE model, due to the remote possibility of an ultra-deep TDE from an off-centre intermediate mass black hole, we consider this sufficiently unlikely that we do not pursue the possibility here. It is worth noting that, while traditional, merger-driven GRBs typically have significant offsets (of order a few kpc) from the nucleus of their host galaxy, there is a small number of GRBs that are more consistent with a nuclear location \citep[see, e.g., the sample in][and \citealt{Ravasio19,deUgartePostigo20,Michalowski18,Tanga18} for the specific cases]{Fruchter06} and thus may be of interest in the context of the ultra-deep TDE model. 

While we cannot distinguish between the merger-driven long GRB model of \citetalias{Levan23} and the model we examine here, the stream self-intersection emission offers a chance to do so in future nuclear GRBs. Wide field soft X-ray instruments are ideal for capturing this behaviour, and the rapidly approaching science operations of satellites like \textit{Einstein Probe} and \textit{SVOM} will provide an opportunity to identify new ultra-deep TDE candidates.

\begin{acknowledgments}
We thank EREF for valuable discussion and AJREF for her insight at only six months old. We also thank the anonymous reviewer for their helpful comments.
RAJEF acknowledges support from the UK Space Agency and the European Union’s Horizon 2020 Programme under the AHEAD2020 project (grant agreement number 871158). CJN acknowledges support from the Science and Technology Facilities Council (grant No. ST/Y000544/1) and from the Leverhulme Trust (grant No. RPG-2021-380). ERC acknowledges support from the National Science Foundation through grant AST-2006684.
\end{acknowledgments}

\vspace{5mm}

\software{XSPEC \citep{xspec}}

\clearpage

\bibliography{main}
\bibliographystyle{aasjournal}

\end{document}